\documentclass[twocolumn, nofootinbib, prd, aps, superscriptaddress, amsmath, tightenlines]{revtex4-1}
\pdfoutput=1

\usepackage[utf8]{inputenc}
\usepackage{graphicx}
\usepackage{subfigure}
\usepackage{placeins}
\usepackage{slashed}
\usepackage{verbatim}
\usepackage{amssymb}
\usepackage{amsmath}
\usepackage{makecell}
\usepackage{hyperref}
\usepackage{tabularx}
\usepackage{booktabs}
\usepackage{enumerate}
\usepackage[titletoc]{appendix}
\usepackage{color}
\usepackage{hyperref}

\newcommand{\nicpb}{National Institute of Chemical Physics and Biophysics, Akadeemia tee 23, 12618 Tallinn, Estonia}
\newcommand{\ut}{University of Tartu, Institute of Physics, Ravila 14c, 50411 Tartu, Estonia.}
\newcommand{\trieste}{Universit\`a degli Studi di Trieste, Strada Costiera 11, I-34151 Trieste, Italy \\
and INFN, Sezione di Trieste, Via Valerio 2,  I-34127 Trieste, Italy.}

\newcommand{\LG}{\mathcal{L}}

\newcommand{\bigO}{\mathcal{O}}
\newcommand{\mixing}{\xi}
\newcommand{\milli}{\epsilon}

\newcommand{\td}{\mathrm{d}}

\newcommand{\lrh}{\leftrightharpoons}

\newcommand{\sv}{\left\langle \sigma v_{M{\o}l} \right\rangle}

\newcommand{\gsim}{\lower.7ex\hbox{$\;\stackrel{\textstyle>}{\sim}\;$}}
\newcommand{\lsim}{\lower.7ex\hbox{$\;\stackrel{\textstyle<}{\sim}\;$}}

\newcommand{\GeV}{\mathrm{GeV}}
\newcommand{\TeV}{\mathrm{TeV}}

\begin{document}
\title{Dark matter and spin-1 milli-charged particles}

\author{Emidio Gabrielli}
\affiliation{\trieste}
\affiliation{\nicpb}

\author{Luca Marzola}
\affiliation{\nicpb}
\affiliation{\ut}

\author{Martti Raidal}
\affiliation{\nicpb}
\affiliation{\ut}

\author{Hardi Veerm\"ae}
\affiliation{\nicpb}

\date{\today}

\begin{abstract}
New physics scenarios beyond the Standard Model predict the existence of milli-charged particles. So far, only spin-1/2 and spin-0 milli-charged particles have been considered in literature, leaving out the interesting case of spin-1. We propose a minimal unitary and renormalizable model of massive milli-charged vector particles. Unitarity requires that these particles are gauge bosons of a non-abelian spontaneously broken gauge symmetry. The minimal scenario then consists of an extended Standard Model gauge group $SU(2)_L\times U(1)_Y\times SU(2)_D$ together with a $SU(2)_D$ dark Higgs boson responsible for the symmetry breaking in the dark sector. By imposing that the dark Higgs multiplet has a non-vanishing milli-hypercharge, stable milli-charged spin-1 fields arise thereby providing a potential dark matter candidate. We analyse the phenomenological constraints on this scenario and discuss their implications.
\end{abstract}

\maketitle


\section{Introduction}

Although we observe electric charge quantization in Nature, this property is not a requirement of the Standard Model (SM) \cite{Foot:1990mn}. New physics beyond this framework could enforce the charge quantization, grand unified theories \cite{Georgi:1974sy} for instance, and the existence of magnetic monopoles, if experimentally corroborated,  would demand it \cite{Dirac:1931kp}.

Other theories beyond the SM, however, predict the existence of particles with a small, either non-quantized or effective, electric charge $\milli e$. Given that $\milli \ll 1$ and $e$ is the electron charge, these particles are usually referred to as milli-charged particles (MCP) \cite{Ignatiev:1978xj,Holdom:1985ag,Abel:2003ue,Batell:2005wa}. MCP arise naturally in theories where two or more unbroken $U(1)$ gauge symmetries, coupled to different matter sectors, possess non-diagonal kinetic terms. Even in the absence of a tree-level kinetic mixing, a non-diagonal kinetic term is inevitably induced by radiative corrections \cite{Holdom:1985ag}. 

MCP are stable because of charge conservation and thus are a natural candidate for dark matter. In this regard, their direct coupling to the SM photons also provides suitable production mechanisms as the thermal freeze-out \cite{Kolb:1990vq, Gondolo:1990dk} or freeze-in \cite{Hall:2009bx}. The possibility of vector MCP has not yet been explored in literature, where the dedicated studies focused on the contexts of spin-0 and 1/2 \cite{Feldman:2007wj,Cline:2012is}. However, the case of spin-1 MCP, $V_\mu$, presents and intriguing feature: as a result of the interplay between gauge interactions and unitarity, the total cross section of $\gamma \gamma \to VV$ tends to a constant in the high energy limit, whereas the same quantity decreases as $s^{-1}$ in the cases of spin-0 and spin-1/2 MCP. Such distinguishing characteristic of spin-1 interactions is manifest in the SM, where the tree-level total cross section for $\gamma\gamma \to W^{+} W^{-}$ approaches a constant of about 80~pb at high energies \cite{Pesic:1973fi,Ginzburg:1982bs,Katuya:1982ga}, while radiative corrections are typically of order 10\% \cite{Denner:1995jv}.

Furthermore, while the interactions of scalars and fermions with the photon can always be induced via kinetic mixing of different $U(1)$ gauge sectors, the spin-1 case requires the extension of the SM gauge group to a larger non-abelian gauge group. 
Motivated by these appealing characteristics, we present here a minimal model of spin-1 milli-charged particles, which preserves both unitary and renormalizability, and investigate its phenomenological implications.  
 
Adopting a minimalistic approach, we consider the $SU(2)_L\times U(1)_Y\times SU(2)_D$ group, where  $SU(2)_D$ is the gauge group of the dark sector that operates on the components $V_{\mu}$ of a dark gauge multiplet. In order to provide the latter with a mass, we also introduce a $SU(2)_D$ Higgs multiplet with a non-zero hypercharge. After symmetry breaking occurred in both the dark and visible sectors, two of the vector fields $V_{\mu}$ acquire masses and milli-charge couplings $\milli$ to the ordinary photon. The remaining vector field also acquires a mass, however it does not couple to the photon. The proposed model then resembles the St\"{u}ckelberg $Z'$ extension of the SM \cite{Kors:2004dx,Kors:2005uz,Cheung:2007ut}, with the important difference that a non-abelian gauge group is at the basis of our construction.

In the absence of $SU(2)_D$ matter fields other than the dark Higgs multiplet, the conservation of electric charge guarantees that our massive vector MCP are stable and, therefore, a potential dark matter candidate. In this regard, the model of spin-1 MCP we present naturally yields a rich and interacting dark sector without introducing a dark or hidden photon \cite{ArkaniHamed:2008qn,Ackerman:mha,Aarssen:2012fx,Tulin:2013teo}. Whereas the phenomenological bounds on the latter clearly do not affect our construction, in the following we will investigate the viability of the proposed scenario by analysing the relevant collider, astrophysical and cosmological constraints.  

The paper is organised as follows. In the next Section we present the most minimal renormalizable model for charged spin-1 fields, in Section~\ref{sec:pheno} we will analyse the phenomenology of this scenario, ranging from astrophysics and cosmology to low-energy experiments. Our conclusions are reported in Section~\ref{sec:conclusions}.

\section{Milli-charged vector bosons and the Standard Model}
\label{sec:model}

\subsection{Lagrangian for charged spin-1 fields}

Charged and massive spin-1 particles can be described by complex vector fields $V_{\mu}$, $\mu \in \{0,1,2,3\}$. Restricting ourself to operators of dimension 4 or lower, the most general, $U(1)_{EM}$ invariant and parity conserving Lagrangian for interacting charged spin-1 fields is~\cite{Ferrara:1992yc,Nieves:1996ff}
\begin{align}
	\label{eq:Lag_VVA}
	\LG_{EM + V}
	=	
&	-\frac{1}{4}F_{\,\mu\nu}F^{\mu\nu}
	- \frac{1}{2} V_{\mu\nu}V^{\dagger\mu\nu}
	+ m_{V}^{2}|V|^{2} 
	\nonumber\\
&	- iQ_{V} (g-1) F_{\,}^{\mu\nu}V_{\mu} V^{\dagger}_{\nu}
	\nonumber\\
&	- \frac{Q_{V}^{2}}{2} (\lambda_{1}(|V|^{2})^{2} - \lambda_{2}|V^{2}|^{2}),
\end{align}
where $Q_{V}$ is the electric charge of the particles associated to $V_\mu$, $g$ their gyromagnetic ratio and $m_V$ their mass. With $F_{\mu\nu}\equiv \partial_{[\mu}A_{\nu]}$ we denote\footnote{The bracket notation indicates here $A_{[\mu} B_{\nu]} := A_{\mu} B_{\nu} - A_{\nu} B_{\mu}$.} the usual  field strength of electromagnetism, while $V_{\mu\nu}$ is the field tensor of the charged vector fields $V_{\mu}$ after imposing the minimal substitution: $V_{\mu\nu} \equiv (\partial + iQ_{V} A)_{[\mu}V_{\nu]}$. The parameters $\lambda_{1}$ and $\lambda_{2}$ govern the self interactions of the fields $V_\mu$. 

We remark that for fields of spin higher than 1/2, the interactions with the photons are not uniquely determined by the minimal substitution, because of the non-commutativity of the covariant derivatives. Requiring the unitarity of the theory, however, further reduces the set of free parameters to the mass $m_{V}$ and charge $Q_{V}$. To be precise, the unitarity of $\gamma \gamma \to VV$ scattering sets the gyromagnetic ratio to $g=2$, while the unitarity of $VV \to VV$ necessarily yields $\lambda_{1} = \lambda_{2} = 1$ \cite{Ferrara:1992yc}.

In spite of that, the presence of an explicit mass term still leads to unitarity violations, although in softer way, and spoils the renormalizability of the model. These problems are solved by noticing that, if the mass term was to be excluded from the Lagrangian in Eq.~\eqref{eq:Lag_VVA}, the choice of parameters $\lambda_{1} = \lambda_{2} = 1$ and $g=2$ would enlarge the gauge symmetry of the latter from $U(1)_{EM}$ to $SU(2)$. By generating the mass terms via spontaneous breaking of this enlarged symmetry, $SU(2) \to U(1)_{EM}$ for instance, the unitarity and renormalizability of the theory then follow \cite{'tHooft:1972fi}.
 Notice that the adopted symmetry breaking pattern does not affect the essence of this construction, which relies on the well known result that renormalizable
and unitary interactions between vector fields are gauge interactions 
\cite{LlewellynSmith:1973ey, Cornwall:1974km, Cornwall:1975aq}.

\subsection{A minimal SM extension with milli-charged vector bosons}

In order to accommodate massive milli-charged vector fields within the framework of the SM, we extend the gauge group of the latter to $SU(3) \times SU(2)_{L} \times SU(2)_{D} \times U(1)_{Y}$. The particle content of the theory comprises three new vector fields, $V_\mu$, that transform under the adjoint representation of the dark interaction and a scalar multiplet of $SU(2)_{D}$, $\phi_{D}$, which is responsible for the symmetry breaking in the dark sector and possesses a tiny coupling to the SM $U(1)_{Y}$, i.e. a milli-hypercharge. As the extended symmetry is broken by the Higgs mechanisms in the visible and dark sector to $U(1)_{EM}$, the milli-hypercharge of the dark Higgs $\phi_{D}$ induces opposite electric milli-charges in two of the dark gauge bosons.

The $SU(2)_{D}$ Yang-Mills term for the dark gauge sector is
\begin{align}\label{eq:Lag_SU2_D}
	\LG_{SU(2)_{D}}
&	=	-\frac{1}{4}V_{3\,\mu\nu}V_{3}^{\mu\nu}
	- \frac{1}{2} V_{\mu\nu}V^{\dagger\mu\nu}
	- ig_{D} V_{3\,}^{\mu\nu}V_{\mu} V^{\dagger}_{\nu} 
	\nonumber\\
&	- \frac{g^{2}_{D}}{2} ((|V|^{2})^{2} - |V^{2}|^{2}),
\end{align}
where $g_D$ is the dark gauge coupling and $V_{3\,\mu\nu} \equiv  \partial_{[\mu} V_{3\nu]}$ and $V_{\mu\nu} \equiv (\partial + ig_{D} V_{3})_{[\mu}V_{\nu]}$ are respectively the field tensors of the vector bosons $V_{3}$ and $V\equiv (V_1 + i V_{2})/\sqrt{2}$.
 Although the presented expression for the $SU(2)_{D}$ Yang-Mills Lagrangian might seem odd, this form makes explicit that the Lagrangian in Eq.~\eqref{eq:Lag_SU2_D} recovers the one in Eq.~\eqref{eq:Lag_VVA} if the mass term is omitted, $g=2$, $\lambda_1=\lambda_2=1$ and $A$ is replaced by $V_3$. 

As previously pointed out, in order to guarantee the unitarity of the theory the mass term for the above vector bosons has to emerge from a Higgs sector. By denoting with $\phi$ the SM Higgs doublet and $\phi_D$ the dark Higgs of milli-hypercharge $Y_{0}$, we give the corresponding Lagrangian as
\begin{align}
	\LG_{h,h_{D}}
&	=	\left|D_\mu \phi_{D}\right|^{2} 
	- 	m_{\phi_{D}}^{2}|\phi_{D}|^{2}
	- 	\lambda_{\phi\phi_{D}} |\phi|^{2} |\phi_{D}|^{2}
	\nonumber\\
&	+ 	\left|D_\mu \phi \right|^{2} 
	- 	m_{\phi}^{2}|\phi|^{2} 
	- 	\lambda_{\phi}|\phi|^{4}
	- 	\lambda_{\phi_{D}}|\phi_{D}|^{4},
\label{eq:Lag_higgs}
\end{align}
where $D_\mu$ stands for the covariant derivative
\begin{align}\label{eq:CD}
	D_\mu 
&	= \left(\partial  - i g W^{i}\tau_{i} - i g' B Y - i g_{D} V^{a}\tau_{D,a} \right)_{\mu}
\end{align}
and $\tau$, $\tau_D$ are the generators of $SU(2)_{L}$ and $SU(2)_{D}$ respectively. The extended gauge group is spontaneously broken once $\phi$ and $\phi_D$ acquire vacuum expectation values $v$ and $v_D$, respectively. The quantum numbers of the vacuum state of the theory are listed in Table~\ref{tab:higgses}.

\subsection{Milli-charge from mass mixing}

\begin{table}[t]
	\centering
\newcolumntype{Y}{>{\centering\arraybackslash}X}
\begin{tabularx}{0.48\textwidth}{YYYYY}
\hline\hline
			& $I_3$	& $I_{3D}$	& $Y$		& Q	\\ 
\hline
$h$		& -1/2	& 0			& 1/2		& 0	\\
$h_{D}$	& 0	 	& $I_{3D,0}$& $Y_{0}$ 	& 0 \\ 
\hline\hline
\end{tabularx}
\caption{The quantum numbers of vacuum in the considered SM extension.}
\label{tab:higgses}
\end{table}

The electric charge of a particle species can be deduced from the following two basic facts: first, a residual $U(1)_{EM}$, or equivalently a massless photon, is possible only if the vacuum is electromagnetically neutral, and second, the electric charge $Q$ of any particle species can be expressed as a linear combination of the hypercharge $Y$, the weak isospin $I_{3}$ and the dark isospin $I_{3D}$. Given the Higgs charge and isospin assignments listed in Table~\ref{tab:higgses}, setting the electron charge to $-e$ uniquely fixes the charges of the remaining particle content according to the generalised Gell-Mann–Nishijima formula
\begin{align}\label{eq:general_Q}
	Q = e(I_{3} + \milli I_{3D} + Y),
\end{align}
where $\milli \equiv -Y_0/I_{3D,0}$. It is then evident that every particle with a non-vanishing dark isospin carries also an electric charge. In particular, the three dark gauge bosons with dark hypercharges $I_{3D} = 0, \pm 1$ result after the symmetry breaking into a neutral $Z'$ boson and two vector bosons $V,$ $V^{\dagger}$, with an electric charge $\milli e$.

Let us remark on the structure of the adopted scheme. The most common alternative for generating MCP is via a kinetic mixing of two or more $U(1)$ gauge fields. Applying such a scheme to the spin-1 case, however, necessarily requires that the SM gauge group be extended at least to the $SU(2)_{D} \times U'(1)$ group\footnote{Notice that, up to fields redefinition, 
imposing a small tree-level kinetic mixing among two 
$U(1)\times U(1)^{\prime}$ gauge fields 
is equivalent to the direct assignment of an effective milli-charge coupling 
\cite{Holdom:1985ag}}. The proposed model of vector MCP arising from the mass mixing is, therefore,  the most minimal choice in terms of gauge and scalar fields, containing only renormalizable interactions.

\subsection{The neutral gauge bosons}

Spontaneous symmetry breaking generates the following mass terms for the gauge bosons 
\begin{align}
	\label{eq:Lag_mass}
	\LG_{M}
&	=	m_V^{2}V_{+}V_{-}
	+	m_W^{2}W^{+}W^{-} 
	+	\frac{1}{2}E^{T} M^{2} E,
\end{align}
where $m_W$ and $m_V$ are the masses of the SM and dark sector charged gauge bosons respectively. The masses of the neutral gauge bosons are instead contained in the matrix $M^{2}$, which in the basis $E^{T} = (V_3,B,W_3)$ is written as
\begin{align}\label{eq:big_mass_matrix}
	M^{2}	=
\begin{pmatrix}
	m_{Z'_{0}}^{2} 			& -m_{Z'_{0}}^{2}\mixing							& 0	\\
	-m_{Z'_{0}}^{2}\mixing  & s_{W}^{2}m_{Z_{0}}^{2} + m_{Z'_{0}}^{2}\mixing^{2}& -m_{Z_{0}}^{2}c_{W}s_{W}	\\
	0						& -m_{Z_{0}}^{2}c_{W}s_{W}							& m_{Z_{0}}^{2}c_{W}^{2}
\end{pmatrix}.
\end{align}
Here we introduced the small dimensionless expansion parameter 
\begin{align}
	\mixing \equiv \milli\,g'/g_D,
\end{align}
to quantify the deviations from the decoupling limit $\milli=\mixing=0$ in which the SM is recovered. 
Notice that the mass matrix in Eq.~\eqref{eq:big_mass_matrix} matches the one of the St\"{u}ckelberg extension of SM, while the expansion parameter $\mixing$ coincides with the mass ratio $M_1/M_2$ commonly used in the framework of St\"{u}ckelberg $Z'$ models \cite{Kors:2005uz}. In this study we assume $\milli \ll \mixing \ll 1$, as suggested by the phenomenological constraints that the searches of MCP and $Z'$ impose \cite{Jaeckel:2010ni}.

The mass parameters in Eq.s~\eqref{eq:Lag_mass} and~\eqref{eq:big_mass_matrix}
\begin{subequations}
\begin{align}
	\label{eq:mz00000}
	m_{W} &= \dfrac{vg}{2},
	\qquad\qquad\qquad
	m_{Z_{0}} = m_{W}/c_{W},	
	\\
	m_{V} &= v_{D}g_{D} |I_{D3,0}| r, 
	\qquad
	m_{Z'_{0}} = m_{V}/r,		
\end{align}
\end{subequations}
correspond to the tree-level masses in the decoupling limit. Here $s_{W}$, $c_{W} \equiv g/\sqrt{g^{2} + {g'}^{2}}$ denote the sine and the cosine of the SM weak mixing angle $\theta_W$ respectively. The parameter 
\begin{align}
	\label{eq:rpar}
	r \equiv \sqrt{(I_{D,0}(I_{D,0} + 1) - I_{D3,0}^{2})/(2 I_{D3,0}^{2})},
\end{align}
quantifies the effect of the dark Higgs vacuum state and depends on the representation adopted for the latter. In the rest of the paper we consider a $SU(2)_{D}$  dark Higgs doublet, corresponding to the most minimal choice, for which $r = 1$ and therefore $m_{Z'_{0}} = m_{V}$. 
The tree level neutral gauge boson masses are the roots of the polynomial $\det(s - M^{2})$, given in Eq.~\eqref{SU2D_detM} of Appendix~\ref{app:amp_and_cs}, with the zero mass eigenstate corresponding to the photon. The remaining non-vanishing eigenvalues 
\begin{align}
	m_{Z}^{2}
&	= m_{Z_{0}}^2 + \mixing^2 m_{Z'_{0}}^2\frac{m_{Z_{0}}^{2} s_{W}^2}{m_{Z_{0}}^2-m_{Z'_{0}}^2} + \bigO(\mixing^4),
	\\
	\label{eq:mzprimemv}
	m_{Z'}^{2}
&	= m_{Z'_{0}}^2 + \mixing^2 m_{Z'_{0}}^2\left(1 - \frac{m_{Z_{0}}^2 s_{W}^2}{m_{Z_{0}}^2-m_{Z'_{0}}^2}\right) + \bigO(\mixing^4).
\end{align}
give the $Z$ and $Z'$ boson masses. The simplest approach to identify the corresponding eigenvector fields in a perturbative fashion is to start from the basis 
\begin{align}\label{eq:mix_matrix}
\begin{pmatrix}
	Z'_{0} \\ A_{0} \\ Z_{0} 						
\end{pmatrix}	
=
\begin{pmatrix}
	1	& 0		& 0		\\
	0	& c_{W}	& s_{W}	\\
	0	& -s_{W}& c_{W}
\end{pmatrix}
\begin{pmatrix}
	V_{3} \\ B \\ W_{3} 						
\end{pmatrix},
\end{align}
formed by the neutral gauge boson mass eigenstates in the decoupling limit. The mass eigenstates of the mass matrix in Eq.~\eqref{eq:big_mass_matrix} are then given through the rotation
\begin{align}
	\label{eq:mixeig}
\begin{pmatrix}
	Z' \\ A \\ Z 						
\end{pmatrix}		
=
\begin{pmatrix}
	c_{1}c_{2}	& - s_{1}c_{2}	& - s_{2}	\\
	s_{1}		& c_{1}			& 0			\\
	c_{1}s_{2}	& - s_{1}s_{2}	& c_{2}
\end{pmatrix}
\begin{pmatrix}
	Z'_{0} \\ A_{0} \\ Z_{0} 						
\end{pmatrix}	,
\end{align}
where $c_{i}$ and $s_{i}$ stand for the cosines and sines of the mixing angles. In the small mixing approximation, $\mixing\ll1$, we have
\begin{align}
	\theta_{1} 
	\approx -\mixing c_{W},
	\qquad
	\theta_{2}
	\approx -\mixing \frac{m_{Z'_{0}}^{2} s_{W}}{m_{Z'_{0}}^{2}-m_{Z_{0}}^{2}}.
\end{align}
Notice that the photon field comprises here an additional contribution from $Z'_0$ but not from $Z_0$, the $Z$ boson in the decoupling limit. As a consequence the elementary charge in \eqref{eq:general_Q} is modified as
\begin{align}
	e 
	= \left({g'}^{-2}(1 + \mixing^{2}) + {g}^{-2}\right)^{-1/2}
	\approx e_{0} -  e_{0}^{3}\mixing^{2}/2,
\end{align}
where $e_0 \equiv e_{\mixing = 0} = g' c_{W}$. We remark that the leading contribution to the charge and masses is of the second order in $\mixing$, whereas the mixing in Eq.~\eqref{eq:mixeig} contains also first order corrections in this expansion parameter. 

Given the above relations, we choose to parametrise our scenario with the following physical quantities: $m_{W}$, $m_{V}$, $\alpha$, $\milli$, $\mixing$. These will serve as input parameters in the computations of the observables connected to the phenomenology of vector MCP discussed below, where the $r$ parameter in Eq.~\eqref{eq:rpar} is set to $r=1$ as previously explained.
Throughout the following we will also approximate most of the quantities with the lowest order of their $\mixing$ expansion. The subscript that denotes quantities in the decoupling limit can therefore be omitted, with the understanding that the committed error is of higher order in $\mixing$. 

To summarise our results so far, we identified a new mechanism that yields massive vector MCP from a non-Abelian extension of the SM. Because of charge conservation, our MCP are stable and therefore provide a suitable dark matter candidate. Our construction is based on the presence of a dark Higgs field, required by the unitarity and renormalizability of the model, that transforms non-trivially under the gauged symmetry of the dark sector. Provided the dark Higgs has a non vanishing milli-hypercharge, the symmetry breaking dynamics then result in a dark sector composed of a neutral vector boson, $Z'$, and two further spin-1 particles with opposite milli-charge, $V^\pm$. The same dynamics also predict modifications to the Electroweak sector of the SM, which however are negligible at the current level of precision achieved by the dedicated experiments. In order to show that the proposed framework is consistent with dark matter, we now turn our attention to the phenomenology of MCP.

\section{Phenomenology}
\label{sec:pheno}

Although MCP can easily avoid the bound from direct detection and collider experiments owing to their small coupling with the photons, in the past decades dedicated experiments as well as astrophysical and cosmological observations \cite{Prinz:1998ua,Dobroliubov:1989mr,Mohapatra:1990vq,Davidson:1991si,Mohapatra:1991as,Davidson:1993sj,Davidson:2000hf,Dubovsky:2003yn,Gies:2006ca, Melchiorri:2007sq} have severely constrained the available parameter space.

More in detail, for light MCP characterised by a mass scale $m_\epsilon$ well below the electron one, production processes like the pair production in an external static magnetic fields \cite{Tsai:1974fa,Erber:1966vv} exhibit non-perturbative effects that lead to large enhancements in the production rates. The presence of MCP can then be detected by investigating the birefringence and dichroism of polarised laser beams that propagate in a strongly magnetised vacuum \cite{Gies:2006ca}. To date these experiments cast the most severe laboratory constraint on light MCP: $\milli < {\cal O}(10^{-6} - 10^{-4})$. Strong bounds are also brought by the invisible decay of orthopositronium \cite{Mitsui:1993ha} and from the Lamb-shift measurements, which limit the MCP contribution that adds to the QED expectations \cite{Lundeen:1981zz}. On the explored mass range, $10^{-7}\, {\rm eV}\lsim m_{\epsilon} \lsim 10^{5}\, {\rm eV}$, these experiments impose $\milli <{\cal O}(10^{-4})$. From the observational point of view, the production of sub-eV MCP in photon-photon collisions yield distortions in the energy spectrum of the Cosmic Microwave Background radiation (CMB) \cite{Dubovsky:2003yn,Melchiorri:2007sq}. The non detection of such distortions then implies a bounds of the order of $\milli \lsim 10^{-7}$ on the milli-charge.  
Stronger but model dependent bounds on this parameter are also brought by stellar evolution, which constrains $\milli <{\cal O}(10^{-14})$ on the range 
$m_{\epsilon} < {\cal O}(10 \, {\rm KeV})$, and by Big Bang Nucleosynthesis, for which $\milli <{\cal O}(10^{-9})$ if $m_{\epsilon} < {\cal O}(10\, {\rm MeV})$ \cite{Raffelt-book}.

Heavier MCP candidates have been investigated in accelerator experiments,  which resulted in $\epsilon \lesssim 10^{-4},\, 10^{-5}$ respectively for $m_\epsilon \approx {\cal O}(1)$ MeV and $m_\epsilon \approx {\cal O}(100)$ MeV at the 95\% confidence level \cite{Prinz:1998ua}. A current proposal for a new experiment at the LHC, \cite{Haas:2014dda}, aims to constrain MCP with masses $ 0.1 \lesssim m_\epsilon \lesssim 100$ GeV and couplings 
$10^{-3} \lsim \, \milli \, \lsim 10^{-1}$.
On the cosmology side, strong bounds for $m_\epsilon< 10 $ GeV have been derived by analysing the impact of MCP on the angular power spectrum of CMB. MCP that are kinetically coupled to the baryon and electrons at the recombination era take part in the acoustic oscillations of the baryon-photon plasma \cite{Dubovsky:2003yn,Dolgov:2013una}, resulting in contribution to the power spectrum degenerate with that of baryons. Independent measurements of the abundance of the latter, from BBN for instance, then cast a severe upper bound on the abundance of MCP at recombination. In particular, for $\milli > 10^{-6}$, MCP can only constitute a subdominant component of dark matter.
Further studies of the recombination epoch dynamics yield the upper bound $\milli < 10^{-6}$ for MCP with mass $m_\epsilon \approx 1$ GeV, which softens to $\milli < 10^{-4}$ for heavier particles with masses of order $10$ TeV~\cite{McDermott:2010pa}. Whereas such particles could be observed at dark matter direct detection experiments \cite{Akerib:2013tjd}, the corresponding bounds are not directly applicable because our galactic magnetic field tends to expel the MCP from the galactic disk \cite{Chuzhoy:2008zy,McDermott:2010pa}.

In the following we analyse the most crucial of these bounds within the proposed framework of vector MCP.

\subsection{$Z'$ phenomenology}

Our scenario predicts the existence of a $Z'$ boson, which is not 
stable in our framework. The width of the $Z'$ boson is given by
\begin{align}
	\Gamma_{Z'} 
	\approx 	
	\dfrac{81 \alpha \mixing^2 m_{V}}{48c_W^{2}}
	\approx 
	2 \times 10^{-2} \mixing^2\, m_{V},
\end{align}
where we used $m_{Z'} \approx m_V$ that holds barring correction of order $\mixing^2$, see Eq.~\eqref{eq:mzprimemv} and the discussion following Eq.~\eqref{eq:mz00000}.

Our $Z'$ shares many properties with the massive vectorial particle appearing in many $U(1)^{\prime}$ extensions of the SM.
The negative results of experimental $Z'$ searches can then be used to constrain the parameter space of our model.

Notice that, since the mass matrix of Eq.~\eqref{eq:big_mass_matrix} coincides with that of the $Z'$ St\"{u}ckelberg extension of the SM, both the theories yield the same $Z-Z'$ mixing phenomenology. As a consequence, the experimental bounds that constrain the St\"{u}ckelberg $Z'$ model can be directly applied to ours. In particular, by requiring the compatibility of our framework with the electroweak precision tests we obtain\footnote{The ratio $M_1/M_2$ of \cite{Chatrchyan:2012it} coincides with our parameter $\mixing$.} $\mixing \lsim  0.06$ \cite{Chatrchyan:2012it}.

On the other hand, the searches for a St\"{u}ckelberg $Z'$ require that $M_{Z'} > 890(540) \mathrm{GeV}$ for $\mixing = 0.06(0.04)$ \cite{Chatrchyan:2012it}. 
From these results, it follows that less restrictive lower bounds on our $Z'$ mass can be achieved by decreasing the value of the parameter $\mixing$.

\subsection{Dark matter relic abundance}

\begin{table}
	\centering
	\begin{tabularx}{.9\linewidth}{>{\centering}X>{\centering}X}
	\hline\hline
	\textbf{Process} & \textbf{Amplitude order}
	\tabularnewline
	\hline
	$W^+ \, W^-, f^+ \, f^- \lrh  V^+ \, V^- $ 		& $\milli$ 
	\tabularnewline
	$\gamma \, Z', Z \, Z' \lrh V^+ \, V^- $ & $\milli^{2}/\mixing$ 
	\tabularnewline	
	$\gamma \, \gamma, \gamma \, Z, Z \, Z \lrh V^+ \, V^-$ 	& $\milli^{2}$ 
	\tabularnewline
	$Z' \, Z' \lrh V^+ \, V^- $ 	& $\milli^{2}/\mixing^{2}$
	\tabularnewline
	$Z' \lrh f^+ \, f^-, W^+ \, W^-$ 		& $\milli$
	\tabularnewline
	\hline\hline
	\end{tabularx}
	\caption{Processes that regulate the abundance of $V^\pm$ and $Z'$ in the early Universe and the relative amplitudes order, with respect
to their leading dependence by the parameter $\mixing$ and milli-charge $\epsilon$ defined in section II-D.}
	\label{tab:processes}
\end{table}

We investigate now the conditions for our vector MCP to populate the dark sector of the Universe. In order to identify a tentative dark matter candidate amongst the available particles, we first consider the interactions that regulate the abundances of $V^\pm$, $Z'$, and $\phi_D$ in the early Universe. The processes shown in Table~\ref{tab:processes} link these particle species in a way that the evolution of the corresponding abundances can only be tracked by solving a set of coupled Boltzmann equation. Given the purpose of the present paper, we choose to defer such a detailed analysis to a dedicated work \cite{in-preparation} 
and propose here a simpler scenario that nevertheless demonstrates the viability of the model. The simplifying assumptions that we consider are
\begin{enumerate}[i)]
	\item $\milli\ll 1$, $\mixing\ll 1$, hence we disregard the processes with amplitudes of higher order in $\milli$ and $\mixing$.
	\item $m_{\phi_D} > 2 m_{Z'}$, barring $\phi_D$ as a dark matter candidate because of the $SU(2)_D$ interactions.
	\item The $\phi_D$--$\phi_{SM}$ mixing is negligible.
	\item 
		  $\milli \ll \mixing^{2}$ , barring the $Z'$ as a possible dark matter candidate and effectively decoupling the dynamics of $Z'$ and $V^\pm$ (see Table~\ref{tab:processes}).
\end{enumerate}   

Under the mentioned conditions $V^\pm$ emerges as the only viable dark matter candidate. The stability of these particles is insured by the conservation of the electric charge, in absence of lighter, coupled, fermion fields of charge $\milli$. Barring the production via scalar mixing, the abundance of $V^\pm$ is modified only by the s-channel pair-production/annihilation from and into the SM charged particles and photons. A direct calculation, reported in Appendix~\ref{app:amp_and_cs}, reveals that the cross section for the $V^\pm$ annihilation into fermions is always larger than the one for annihilation into $W^\pm$ by about two order of magnitude. We will then disregard the impact of the latter in our computations.  The $V^\pm$ abundance is regulated by the Boltzmann equation \cite{Kolb:1990vq,Gondolo:1990dk}
\begin{equation}\label{eq:boltz1}
	\dot{n}_{V} +3 H n_{V} 
	= 
	-\frac{1}{2}\sv \left(n_{V}^2 - n_{V, eq}^2\right),
\end{equation}
where a dot stands for the differentiation with respect to the coordinate time, $\sv$ is the thermally averaged interaction rate and $n_{V, Eq}$ is the equilibrium density of the milli-charged bosons. In the absence of a charge asymmetry the total number density of vector milli-charged particles is $n_V := n_{V^+} + n_{V^-} = 2n_{V^\pm}$. Assuming the conservation of entropy, as well as that the relevant dynamics takes place in the radiation dominated regime, Eq.~\eqref{eq:boltz1} can be recast as
\begin{align}\label{eq:boltz2}
	\dfrac{\td Y}{\td x}
&	=	- \lambda
		\left(Y^{2} - Y_{\,eq}^{2}\right),	
\end{align}
where $Y := n_{V}/s$ is the comoving density normalised to the entropy density $s$, $Y_{eq}:= n_{eq}/s$, $x := m/T$ and we defined
\begin{align}
	\lambda :=  
	g_{*}^{1/2}\sqrt{\frac{\pi}{45}} m_{V} M_{\mathrm{pl}}\frac{1}{2}\langle \sigma \,v_{\text{M{\o}l}}\rangle \,x^{-2}
\end{align}
with the M{\o}ller velocity
\begin{equation}
	v_{M{\o}l} = \frac{\sqrt{(p_1 \cdot p_2) - m_1^2 m_2^2}}{E_1 E_2}
\end{equation} 
and the labels `1' and `2' are referred to the initial state particles.
In the above equation $g_{*}$ quantifies the effective number of relativistic degrees of freedom and $M_{\mathrm{pl}}$ is the Planck mass. The present abundance of spin-1 MCP
is given by
\begin{align}
	\Omega_V h^2 
	\approx 2.8 \times 10^{11} \frac{m_{V}}{\TeV} Y_{0}.
\end{align}
where $Y_0$, the comoving density of MCP at the present time, is obtained by integrating Eq.~\eqref{eq:boltz2} until $x_0 = m_V / T_0 $, being $T_0$ the CMB temperature of today.
\\
The parameters of the proposed model are then clearly constrained by the requirement that the abundance of MCP does not exceed the measured dark matter one: $\Omega_V h^2 \leq \Omega_{DM} h^2 = 0.1199(27)$ \cite{Ade:2013zuv}. In order to calculate $\Omega_V$, we consider two complementary production mechanisms: the thermal freeze-out and freeze-in.

\subsubsection{Production via freeze-out}

In our scenario, the relevant process for the freeze-out mechanism is the $s$-channel annihilation of $V^{^\pm}$ to SM particles. As remarked before, the $t$-channel processes are indeed suppressed by a higher order of the milli-charge. The corresponding cross-sections at the threshold $s \approx 4 m_V^2$ are
\begin{align}\label{eq:sigma_limits_VVtoSM_TH}
	\sigma_{V^{+}V^{-} \to SM} 
&	=	\dfrac{9\pi \alpha^2}{16 c_W^{4}} \dfrac{\milli^2 }{m_{V}^2}\beta_V
	+	\frac{38\pi \alpha^2}{9} \dfrac{\milli^4 }{m_{V}^2} \beta_V^{-1},
\end{align}
where $\beta_V = \sqrt{1 - 4m_V^{2}/s}$ is the speed of $V^{\pm}$ in the centre of mass frame. The second term corresponds to the leading order contribution from $V^{+}V^{-} \to \gamma\gamma$ that we report for completeness. Notice that although suppressed by a higher power of the milli-charge, this term presents a $1/\beta_V^{2}$ enhancement at low energies with respect to the first one. Hence, although the corresponding contribution is certainly negligible during the freeze out dynamics, it could give rise to observable effects at later eras which we plan to investigate when dealing with the full model. 

As the temperature drops below the mass of the particle, the inverse annihilation rate becomes smaller than the expansion rate of the Universe and Eq.~\eqref{eq:boltz2} reduces to $Y' = - \lambda Y^{2}$. Integrating this equation yields a final abundance which is inversely proportional to the cross-section:
\begin{align}
		Y_{0}^{-1}
	= 	Y_{f}^{-1} + \int^{x_{0}}_{x_f}\td x \, \lambda.
\end{align}
Here $x_f$ is the freeze out temperature, determined through the condition $(Y_{eq}^{-1})'(x_f) = \, \delta(\delta+1) \lambda(x_f)$ where $\delta$ is a numerical constant of order $\bigO(1)$. 

The abundance at the freeze out, $Y_{f} := Y(x_f)$, is typically much larger than the present abundance $Y_0$ and the term $Y_{f}^{-1} $ can then be safely neglected at the desired accuracy order. Then, by approximating the thermally averaged cross section with the leading order of its low temperature expansion
\begin{align}
	\langle \sigma \,v_{\text{M{\o}l}}\rangle 
	\approx \dfrac{27\pi \alpha^2}{16 c_W^{4}} \dfrac{\milli^2 }{m_{V}^2} x^{-1},
\end{align}
we obtain the freeze-out temperature\footnote{In our estimates we take the value of the fine structure constant as given at the electroweak scale.}
\begin{align} 
	x_f  \approx 
	8 + \ln\left(\left(\frac{\milli}{10^{-3}}\right)^{2} \left(\frac{m_V}{\TeV}\right)\right) ,
\end{align}
corresponding to a relic abundance of
\begin{align}\label{eq:fo_omega_mcp}
	\Omega_{MCP} h^2 \approx 2.4 \times 10^{5} x_f^{2} \left(\frac{m_V}{\TeV}\right)^{2}\left(\frac{\milli}{10^{-3}}\right)^{-2}.
\end{align}
Imposing the upper bound $\Omega_{MCP} h^2 < 0.12$ results in the following rough upper bound on the MCP mass in the region $\milli = \bigO(10^{-2} - 10^{-8})$: 
\begin{align}
	m_V \lesssim \milli \times 100 \quad \GeV. 
\end{align}
Given the constraints mentioned at the beginning of the Section, the freeze-out mechanism does not allow for vector MCP dark matter in this simplified scheme:
dark matter candidates in the allowed mass range yield an overabundance that cannot be depleted in this simplified scheme. The necessary higher annihilation cross section could however be achieved by considering other annihilation channels, that could dominate over the proposed milli-charge channel, or by introducing resonances in the latter. The minimal model we propose already contains suitable candidates to implement both the mentioned enhancement mechanisms and, in a future paper \cite{in-preparation}, we intend to study the impact of the Higgs portal and the dark Higgs boson as well as the possible role of the $Z'$ boson in determining the dark matter relic abundance.

For instance, allowing for $\milli \simeq \mixing$ and therefore $g_D \simeq g'$ yields a scenario in which the abundances of $V^\pm$ and $Z'$ are in equilibrium with each other in the early Universe. The decays and inverse decays of the $Z'$ boson would then provide additional thermal contact between the visible and dark sectors and control, along with the reaction $Z' \, Z' \lrh V^+ \, V^- $, the DM freeze-out production.

\subsubsection{Production via freeze-in}

The freeze-out scenario is not successful because of the too large annihilation cross-section that it requires. It is then plausible that the desired spin-1 MCP abundance could instead arise via the freeze-in mechanism \cite{Hall:2009bx}. In this scheme, the particles being produced never achieve the equilibrium density, so that $Y \ll Y_{\,eq}$ and the Boltzmann equation in Eq.~\eqref{eq:boltz2} reduces to\footnote{Given the relation
$\sigma_{ V^{+}V^{-} \to SM } Y_{\,eq}^{2} = \sigma_{ SM -> V^{+}V^{-}} Y_{SM\,eq}^{2}$ 
that holds between the equilibrium densities, we employ the annihilation cross section in Eq.~\eqref{eq:sigma_limits_VVtoSM_HE} to describe the inverse annihilation rate.}  
\begin{equation}
Y' = \lambda Y_{\,eq}^{2}	
\end{equation}
and the MCP production takes place at energies much higher than the corresponding mass scale. 
The MCP equilibrium density is then approximately constant, $Y_{\,eq} \approx 0.28 \, g_{V}/g_{*}$, with $g_{V} = 6$ denoting the degrees of freedom of $V$. The inverse annihilation that drives the MCP production is effectively Bose-enhanced, resulting in a contribution at most of order $\bigO(1)$ that we however neglect in the present analysis. 

For $T> m_V$, the relevant cross-section is approximated as 
\begin{align}\label{eq:sigma_limits_VVtoSM_HE}
	\sigma_{ V^{+}V^{-} \to SM } 
&	=	\dfrac{3\pi \alpha^2}{16 c_W^{4} s} \milli^2 + \dfrac{32\pi \alpha^2}{9m_V^2}\milli^4
\end{align} 
where, on top of the $s$-channel fermion annihilation contribution proportional to $\milli^2$, we considered the $t$-channel processes $\gamma \gamma \to V^+V^-$. The possible relevance of the latter stems from its constant behaviour in the high energy limit: even if this process is of higher order in $\milli$, the corresponding cross section does not present the typical $1/s$ suppression at high energies, therefore this channel dominates the $V^{\pm}$ production as long as the reheating temperature is considerably higher than $m_V$. 

The corresponding thermal average is then given by
\begin{align}
	\langle \sigma \,v_{\text{M{\o}l}}\rangle 
	\approx \frac{\pi \alpha^2}{m_V^2}\left(\frac{32}{9} \milli^4 
	+ 		\frac{3}{128 c_w^{4}} \milli^2 x^{2}\right),
\end{align}
and by supposing that the MCP have a vanishing initial abundance and are produced only via the inverse annihilations, we obtain a relic abundance of
\begin{align}
	\Omega h^2  
	\approx	2.1 \times 10^{18}
	\left(\milli^2 x_{s} + 100 \, \milli^4 x_{r}^{-1} \right),
\end{align}
where $T_{r} := m_V/x_{r}$ is the reheating temperature after Inflation and $T_s := m_V/x_{s}$ is the effective temperature at which the production of vector MCP stops. 
The above result holds under the assumption that $T_r \gg T_s$ and numerical solutions of Eq.~\eqref{eq:boltz2} give the value $x_{s} \approx 1.36$. The requirement $\Omega h^2 < 0.12$ then leads to the following minimal upper bound on the milli-charge:
\begin{align}
	\milli \lesssim \min(2 \times 10^{-10}, 5\times 10^{-6} x_{r}^{1/4}).
\end{align}
The $x_r$ independent bound, the first term in Eq.~\eqref{eq:sigma_limits_VVtoSM_HE}, holds for MCP of any spin within an accuracy of order $\bigO(1)$. The second bound, instead, is due to the characteristic behaviour of the spin-1$\to$spin-1 cross section. Given that the constraints on the St\"{u}ckelberg $Z'$ set $m_V \gtrsim 0.5\, \TeV$, it follows that the reheating temperature required to create a DM abundance of vector-MCP via this channel is of the order of Planck mass and therefore this channel is currently disfavoured by the CMB observations.

\subsection{Astrophysical plasma effects}

The proposed milli-charged dark matter forms a pair plasma which can affect the dynamics of galaxy cluster collisions by forming shockwaves even if the plasma is effectively collision-less \cite{Heikinheimo:2015kra}. Constraints for dark matter self-interactions have been derived from observations of cluster collisions~\cite{Markevitch:2003at,Harvey:2015hha}. The cleanest example, and  probably the most constraining, is the Bullet Cluster for which it was shown that no more than 30\% of the total DM mass can be stripped from the sub-cluster as it passes through the main cluster halo \cite{Markevitch:2003at}. 

The dynamics of non-relativistic collisions of pair plasmas in the non-linear regime, where we expect a fully formed shockwave, are still not well understood. On the contrary, in the linear regime it is possible to estimate the growth rate of plasma instabilities and therefore also roughly the time scale of shock shock formation \cite{Heikinheimo:2015kra,Bret:2009fy}
\begin{align}
	\tau 
	\approx 10^{3}\omega_{p}^{-1}
	\approx 0.1 \,\mathrm{yr}  \times \left(\frac{m_V}{\TeV}\right) \left(\frac{10^{-3}}{\milli}\right),
\end{align}
where $\omega_{p} = \milli \left(4 \pi \alpha n/m_V\right)^{1/2}$ is the plasma frequency, $n$ denotes the number density of the plasma constituents and we assumed a DM density of $\rho_{DM} \approx 0.1\,\GeV\,\mathrm{cm}^{-3}$. The time scale of a galaxy cluster collision is of the order of $0.1\,\mathrm{Gyr}$. If we require that no shockwaves are formed, then all dark matter may be milli-charged if
\begin{align}
	\milli \lesssim 10^{-9} \left(\frac{m_V}{\TeV}\right).
\end{align}
We stress that this inequality is indicative, at best, and more accurate methods need to be used for deriving a reliable estimate. However, the above result seems to indicate that the shock behaviour of the plasma can be completely neglected in freeze-in scenarios with heavy dark matter $m_V \gtrsim \TeV$. If the inequality was violated as in the presented freeze-out case, then we would expect that astrophysical observations could disprove the vector MCP DM scenario.

Finally we remark that milli-charged dark matter also interacts with the intra-galactic visible plasma. This might have a non negligible impact on the distribution of the X-ray emitting astrophysical plasmas, observed, for example, in cluster collisions. A detailed examination of these effects is however beyond the purpose of the current study.

\subsection{Acoustic plasma oscillations at the recombination}
\begin{figure}[h]
  \centering
    \includegraphics[width=.35\textwidth]{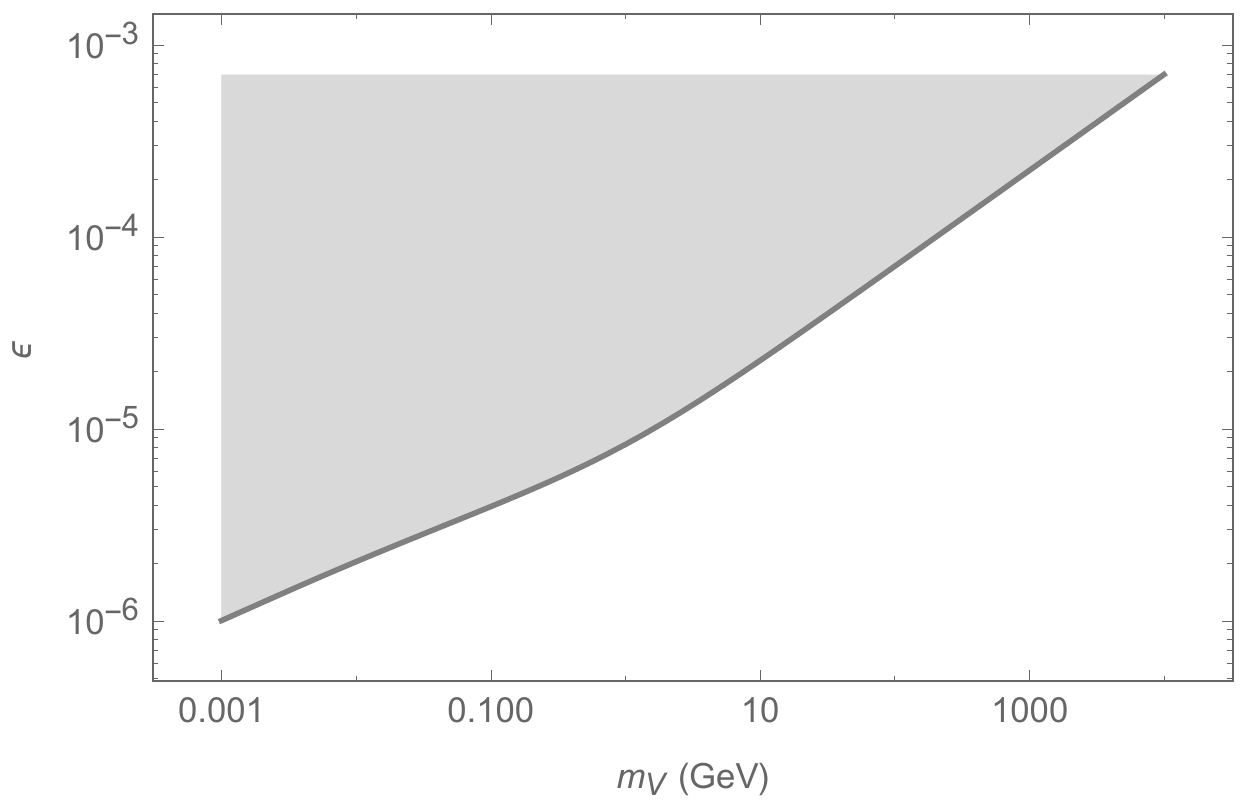}
  \caption{The exclusion area cut in the MCP parameter space by the requirement that MCP do not take part in the acoustic oscillations of the particle plasma at the recombination era, \cite{Dubovsky:2003yn,Dolgov:2013una,McDermott:2010pa}. In our model, the region $m_V \lesssim 500$ GeV is excluded by the lower bound on the $Z'$ mass $m_{Z'} = m_V$ resulting from the dedicated experimental searches \cite{Chatrchyan:2012it}.}
  \label{fig:cmb_bound}
\end{figure}

As previously mentioned, MCP that are tightly coupled to the SM particles at recombination participate in the acoustic oscillation of the particle plasma at recombination, yielding a contribution to the CMB spectrum that is degenerate with the one brought by baryons. Independent measurements of the abundance of the latter then allow MCP to be only a sub-dominant component in the observed dark matter relic density.   
The tight-coupling condition, that forbids dark matter MCP, is quantified in the following inequality \cite{Dubovsky:2003yn,Dolgov:2013una}
\begin{equation}
	\epsilon^2 > 5 \times 10^{-11}\frac{m_V}{\sqrt{\mu_p} + \sqrt{m_e}}\quad\text{GeV}^{-1/2},
\end{equation}
where $\mu_x := m_V m_x/(m_V + m_x)$. The resulting bound on MCP models is reported in Figure~\ref{fig:cmb_bound}, where the grey area denotes the region of the MCP parameter space in which the interactions between MCP protons and electrons are substantial enough to ensure the tight coupling of the former to the latter. Given the constraint cast by $Z'$ searches, this bound is clearly consistent with the candidate we propose.  
In the above derivation, the interaction between photon and MCP have been neglected since the relevant diagrams are of the fourth order in $\milli$.

\subsection{Halo stability}

If dark matter has significant self interactions and a suitable mass spectrum,   the dark matter halos can collapse and reduce to dark disks via cooling through bremsstrahlung or Compton scattering on CMB photons. Due to the absence of a light dark photon, the feeble interactions and the mass scale of the MCP dark matter candidates we are proposing, we expect such a cooling process to be negligible in the present case. To show this, we assume the dark plasma is virialized and has a virial temperature
\begin{align}
	T_{vir} = \frac{M m_V}{n_{dof}M_{pl}^{2}R_{vir}},
\end{align}
where $n_{dof} = 3$ is the number of degrees of freedom carried by a single particle and $M$ and $R_{vir}$ denote the mass and the radius of the virial cluster respectively. The characteristic timescale for dark bremsstrahlung cooling in our model is larger than the age of the Universe by many orders of magnitude \cite{Fan:2013yva}
\begin{equation}
	t_{\mathrm{brems}} 
	\approx \frac{3}{16} \; \frac{m_V^{5/2} T_{vir}^{1/2}}{\alpha^3\rho_V\, \milli^{6}}
	\approx 10^{22} \mathrm{yr} \, \left(\frac{m_V}{\TeV}\right)^{3}\,\milli^{-6},
\end{equation}
where we assumed $T_{vir} \approx 10^{-6} m_V$ and $\rho_{V} \approx 0.1\,\GeV\,\mathrm{cm}^{-3}$ in order to obtain a conservative estimate.  As cooling through Compton scattering requires even larger timescales
\begin{equation}
	t_{\mathrm{Compton}} 
	\approx \frac{135}{64\pi^3} \; \frac{m_V^{3}}{\alpha^2\,T_{\mathrm{CMB}}^{4}\, \milli^{4}}
	\approx 10^{32} \mathrm{yr} \, \left(\frac{m_V}{\TeV}\right)^{3}\,\milli^{-4},
\end{equation}
we conclude that DM haloes composed of our spin-1 MCP do not collapse to disks for the considered values of the parameters.

\section{Summary} 
\label{sec:conclusions}
We proposed the most minimal renormalizable model of spin-1 MCP, based on the SM gauge group extension $SU(2)_L\times U(1)_Y\times SU(2)_D$, containing only a doublet Higgs field in the dark sector which is milli-charged under the $U(1)_Y$ SM gauge group. We studied the basic phenomenological constraints of this model with the stress on a having a viable vectorial milli-charged (component) of dark matter.

Constraints on the milli-charge $\milli e$ versus the mass $m_V$ were analysed by requiring that the vectorial MCP satisfies the constraints imposed by the observed dark matter relic abundance. The production of the latter by both freeze-out and freeze-in has been investigated. In order to propose a first simplified scenario based on the proposed framework, we worked under a simplifying assumption that bars the effects of both the dark Higgs boson and the $Z^{\prime}$ component of $SU(2)_D$ in determining the relic dark matter abundance. 
\\
In this simplified scheme, we found that the freeze-out mechanism yields an overabundance of vector MCP with respect to the measured dark matter abundance. We believe that relaxing our working assumptions could help to achieve the desired relic abundance within the freeze-out scenario and briefly commented on this possibility which we intend to investigate in a dedicated follow-up paper.
\\
On the other hand, within the freeze-in scenario, matching the dark matter relic abundance imposes $\milli \lesssim 10^{-10}$ regardless of the MCP spin.  Owing to the peculiar behaviour of the $\gamma\gamma \to VV$ cross section at high energy, spin-1 MCP present an additional production channel. In this case reproducing the dark matter abundance and respecting the mass constraints imposed by $Z'$ searches however requires a reheating temperature of the order of Planck mass.
\\ 
For the values of mass and milli-charge emerging from the analysed freeze-in scenario, our MCP dark matter candidate do not participate in the acoustic plasma oscillations at the recombination era and, therefore, avoid the severe bound cast by CMB analyses. In a similar fashion, the constraints emerging from dark matter halo stability are met owing to the smallness of the milli-charge $\milli\approx 10^{-10}$ required by the freeze-in mechanism. This value is comparable with the condition for shockwaves formation in cluster collisions that plasma physics indicates, possibly allowing for a test of the scenario owing to the implied astrophysical effects.

\section*{Acknowledgments} We thank D. Comelli, K. Ehat\"{a}ht and E. Milotti for useful discussions.
LM acknowledges the European Social Fund for supporting his work with the MOBILITAS grant MJD387.
This work was supported by grants  MTT60, IUT23-6, CERN+, and by the EU through the ERDF CoE program. EG, LM, and HV would also like to thank the TH-PH division for the kind hospitality during the preparation of this work.
\appendix

\section{Cross sections and decay widths} 
\label{app:amp_and_cs}

Here we report the relevant cross-sections at the leading order in $\mixing$. The cross-sections for $f\bar{f} \to V^{+}V^{-}$ is
 \begin{widetext}
\begin{align}
	\sigma_{f\bar{f} \to V^{+}V^{-}}
&	= 	\frac{\pi \alpha^2 \milli^2}{24}
		\beta_V^{3}\beta_f^{-1}
		\,\det(s-M)^{-2}
		\left(12 m_{V}^4+20 m_{V}^2 s+s^2\right) 
		\\ \nonumber
&       \times\bigg(	
			4 m_{f}^2 m_{W}^4 Y_R^2 
			+ s (2 m_{W}^4 Y_R^2 - 4 m_{f}^2 m_{W}^2 Y_R (Y_L+Y_R)) 
			- \\ \nonumber
			&- s^2 \big( m_{f}^2 (Y_L^2 - 6 Y_L Y_R + Y_R^2) 
		+ 2 m_{W}^2 Y_R (Y_L+Y_R)\big) 
		+	s^3 (Y_L^2+Y_R^2)
		\bigg),
\end{align}
where $\beta_X = \sqrt{1 - 4m_X^{2}/s}$ and $Y_L$, $Y_R$ denote the hypercharge of the left- and right-handed fermion correspondingly. The determinant entering the propagators contains the mass matrix $M^2$, explicitly 
\begin{align}\label{SU2D_detM}
	\det(s - M^{2}) 
	=	s \big((s - m_{Z_{0}}^{2})(s - m_{Z'_{0}}^{2}) - \mixing^{2} m_{Z'_{0}}^{2}(s -  m_{W}^{2})\big)
	\equiv	s (s - m_Z^{2})(s - m_{Z'}^{2}).
\end{align}
The cross-sections for $W^{+}W^{-} \to V^{+}V^{-}$ and $\gamma\gamma \to V^{+}V^{-}$ are
\begin{align}
  	\sigma_{W^{+}W^{-} \to V^{+}V^{-}}
&	=	\frac{\pi \alpha^2 \milli^2}{432 c_w^{4}}
		\beta_V^{3}\beta_W^{-1}\det(s-M)^{-2}
		\times	s
		\left(12 m_{V}^4+20 m_{V}^2 s+s^2\right)\\ \nonumber
		& \times 
		\left(12 m_{W}^4+20 m_{W}^2 s+s^2\right),
	\\
	\sigma_{\gamma\gamma \to V^{+}V^{-}}  
&	= \frac{8\pi \alpha^2 \milli^4}{m_V^2} \beta_{V}
	\bigg(
		\frac{3}{16}(1-\beta_{V}^2) 
		\times	\left(2-\beta_{V}^2 - (1-\beta_{V}^4) \beta_{V}^{-1}\mathrm{atanh}(\beta_{V})\right) 
		+ 	1
	\bigg).
\end{align}
\end{widetext}

In the following we assume that $m_W, m_f \ll m_V$ and thereby neglect the SM masses. The high energy asymptotic of the cross-sections reads
\begin{align}
	\sigma_{W^{+}W^{-} \to V^{+}V^{-}} 
&	= \frac{\pi \alpha^2 \milli^2}{432 c_w^{4} s},
	\\
	\sigma_{f\bar{f} \to V^{+}V^{-}} 
&	= \frac{\pi \alpha^2 \milli^2}{432 c_w^{4} s} \times 18(Y_L^2+Y_R^2) ,
	\\
	\sigma_{\gamma\gamma \to V^{+}V^{-}} 
&	= \frac{8\pi \alpha^2 \milli^4}{m_V^2} ,
\end{align}
and at the threshold $s \approx 4m_V^{2}$ 
\begin{align}\label{eq:sigmath_WWtoVV}
	\sigma_{W^{+}W^{-} \to V^{+}V^{-}} 
&	= 	\frac{\pi \alpha^2 \milli^2}{144 c_w^{4}} 
		\frac{\beta_V^{3}}{m_{V}^2} 	,	
   \\ \label{eq:sigmath_fftoVV}
  	\sigma_{f\bar{f} \to V^{+}V^{-}}
&	= 	\frac{\pi \alpha^2 \milli^2}{144 c_w^{4}} 
		\frac{\beta_V^{3}}{m_{V}^2}
		\times 18 \left(Y_{L}^2+Y_{R}^2\right),
	\\
	\sigma_{\gamma\gamma \to V^{+}V^{-}} 
&	= \frac{19 \pi \alpha^2 \milli^4}{2 m_V^2} \beta_V.
\end{align}

The asymptotics for the $VV$ annihilations into the SM particles are
\begin{align}\label{eq:sigma_limits_VVtoSM}
	\sigma_{V^{+}V^{-} \to SM} 
&	=	
\left\{
\begin{array}{ll}
	 \dfrac{32 \pi \alpha^2 \milli^4}{9 m_V^2} \beta_V^{-1} + \dfrac{9 \pi \alpha^2 \milli^2}{16 c_{w}^{4} m_{V}^2} \beta_V + \bigO(\beta_V^{2}), \\
	 \dfrac{32 \pi \alpha^2 \milli^4}{9 m_V^2} + \dfrac{3 \pi \alpha^2 \milli^2}{16 c_{w}^{4}}s^{-1} + \bigO(s^{-2}).
\end{array}
\right.
\end{align}
Although the $t$-channel or $\milli^4$ term is of higher order in the milli-charge, it dominates the asymptotic regime with large $s$ or small $\beta_V$.

The decay rate of the $Z'$ (assuming $m_{Z'} \gg m_f,m_W$) to SM particles is
\begin{subequations}
\begin{align}
	\Gamma_{Z'\to f\bar{f}} 
&	= 	\frac{\mixing^2 \alpha m_{V}}{48 c_{w}^{2}}\times 8(Y_{R}^{2} + Y_{L}^{2}),
	\\
	\Gamma_{Z' \to W^{+}W^{-}} 
&	= 	\frac{\mixing^2 \alpha m_{V}}{48 c_{w}^{2}}.
\end{align}
\end{subequations}
Assuming a heavy Higgs and $2m_V > m_{Z'}$ these are also the only decay channels. The $Z'$ width is then
\begin{align}
	\Gamma_{Z'} = \dfrac{81 \alpha}{48c_{w}^{2}}\mixing^2 m_{V}.
\end{align}

\section{The Feynman rules for the milli-charge expansion}

In the perturbative prescription, with the mixing as expansion parameter, the SM Feynman rules remain unchanged while there are additional 2-leg vertices from mass mixing connecting the dark and visible sector:
\begin{align}
\begin{array}{lc}
	- i \mixing m_{Z'}^{2} c_{w}		&	\qquad  Z'-\gamma		,\\
	+ i \mixing m_{Z'}^{2} s_{w}		&	\qquad	Z'-Z	,		\\
	+ i \mixing^{2} m_{Z'}^{2}c_{w}^{2}	&	\qquad	\gamma-\gamma	,\\
	- i \mixing^{2} m_{Z'}^{2}c_{w}s_w	&	\qquad	Z-\gamma		,\\
	+ i \mixing^{2} m_{Z'}^{2}c_{w}^{2}	&	\qquad	Z-Z			.	\\
\end{array}
\end{align}
Besides a possible Higgs portal, the above interactions are the only ones that connect the two sectors at a diagram level. The dark sector Feynman rules are the usual rules for a spontaneously broken SU(2).

%

\bibliography{bib}
\bibliographystyle{hunsrt}

\end{document}